\documentclass[%
 reprint,
 amsmath,amssymb,
 aps,
]{revtex4-2}

\usepackage{graphicx}% Include figure files
\usepackage{dcolumn}% Align table columns on decimal point
\usepackage{bm}% bold math
\usepackage{comment}
\usepackage{xcolor}
\usepackage[colorlinks=true, linkcolor=blue]{hyperref}
\usepackage{bbm}
\usepackage{slashed}
\usepackage[normalem]{ulem}
\usepackage{mathrsfs}
\usepackage{physics}
\DeclareMathOperator{\sgn}{sgn}

\begin{document}

\preprint{APS/123-QED}

%\title{Topological L\'evy crystal insulator}% Force line breaks with \\
% \title{Localized massless fermions in spacetime-antispacetime domain wall}
% \title{Robust Topology and Tunable Geometry: A Fractional-Dispersion Generalization of the BHZ Model}
\title{Robust Topology and Tunable Geometry in Generalized BHZ Model with Fractional Dispersion}

\author{Coleen Adrianne Panganiban}
 \altaffiliation[Also at ]{National Institute of Physics, University of the Philippines Diliman, Philippines.}%Lines break automatically or can be forced with \\
\author{Kristian Hauser A.~Villegas}%
 \email{kavillegas1@up.edu.ph}
\affiliation{%
 National Institute of Physics, University of the Philippines Diliman, Philippines.\\
}%

\date{\today}% It is always \today, today,

\begin{abstract}
With recent developments in fractional quantum mechanics, we introduce a fractional generalization of the Bernevig-Hughes-Zhang (BHZ) model to investigate the effects of fractional dispersion on band topology and quantum geometry. In the low-energy limit, the model reduces to a fractional Dirac Hamiltonian while preserving momentum-space periodicity, thereby ensuring a compact Brillouin zone and a well-defined topological invariant. We further show that the corresponding real-space tight-binding model can be constructed directly through a Fourier-series transformation, providing a simpler and more general alternative to the methods typically employed for fractional lattice systems. This approach readily extends beyond the generalized BHZ model considered here. Using the Bloch eigenstates, we compute the quantum geometric tensor and analyze its components, namely the Berry curvature and quantum metric. We find that fractional tuning redistributes these quantities throughout the Brillouin zone as the dispersion exponent becomes fractional, leading to pronounced modifications of the local band geometry. In contrast, the Chern number remains invariant, demonstrating the robustness of the global topological phase against fractional deformation. We further argue that this invariance persists for a broad class of fractional models.

\end{abstract}

\maketitle

\section{Introduction}\label{sec:intro}
The derivative operator can be extended to a fractional form in what is known as fractional calculus \cite{Herrmann2014}. This mathematical extension has found applications in various areas of physics, such as fractional quantum mechanics, where the system exhibits a dispersion relation of the form $E \sim k^{\alpha}$, with $\alpha$ not constrained to integer values \cite{Laskin2018}. Remarkably, this exotic behavior has been experimentally realized in several physical systems. For instance, in condensed matter physics, the phonon dispersion at the boundary between two superfluid helium phases has been observed to follow the dispersion relation $E \sim k^{3/2}$ \cite{Watanabe2014}. Furthermore, fractional dispersion has been identified in a variety of other contexts, including networks exhibiting long-range transport \cite{Riascos2015}, nonlocal field theories \cite{Atmann2022}, and optical phenomena in inhomogeneous linear media \cite{Iomin2021}. In the field of optics, specifically, wave evolution in fractional media like L\'evy crystals exhibits exotic behavior \cite{Zakeri2023}, and the fractional Schr\"odinger equation has been utilized to describe gravitational optics \cite{Iomin2021}.

Integer exponents higher than two also play a critical role in multilayer systems; for example, $N$-layer ABC-stacked graphene exhibits a dispersion of $E(k) \sim k^N$ \cite{Zhang2010}. Bilayer graphene has been shown to interpolate between $\alpha=2$ and $\alpha=4$, with the exponent being tunable by an applied perpendicular displacement field \cite{Dong2023}, suggesting fractional dispersion behavior in between. Modifying this fractional exponent has been shown to have significant effects on unconventional discontinuous transitions in isospin systems \cite{Raines2024}. Beyond condensed matter, the framework of fractional quantum mechanics has found applications in astroparticle physics, including entropy-area relations in black hole thermodynamics \cite{Jalalzadeh2021} and the confinement of quarks in quark-antiquark systems \cite{Abu2024}.

For the past few years, there has been increasing research interest in the quantum geometric tensor (QGT) in multiband systems \cite{Liu}. The real part of the QGT, known as the quantum metric, is only beginning to have its physical implications explored \cite{Torma2023}. Recent studies have highlighted its crucial role in observable phenomena, such as influencing the superfluid weight in the superconducting phase of twisted bilayer graphene \cite{Hu2019}. Additional investigations have shown that the quantum metric profoundly impacts collective modes in superconductors \cite{Villegas2021}, enhances electron-phonon coupling \cite{Yu2024}, and is encoded directly into superconducting pairing potentials \cite{Daido2024}. In contrast, the imaginary part of the QGT corresponds to the well-known Berry curvature, which plays a key role in generating Hall conductivity in transport measurements \cite{Ceresoli2006}. Moreover, the integral of the Berry curvature over the Brillouin zone yields the Chern number, a topological invariant used to classify different types of insulators.

A previous work has investigated the effects of fractional dispersion on the QGT using a generalized fractional Dirac Hamiltonian in two dimensions \cite{Biscocho2024}. This study revealed intriguing redistributions of the Berry curvature and quantum metric in momentum space as a function of the parameter $\alpha$ \cite{Biscocho2024}. However, the Hamiltonian employed is a low-energy effective theory, which lacks both upper and lower bounds for the energy. As a result, the momentum space is not compact, meaning it does not correspond to a proper Brillouin zone. Consequently, the Chern number—a topological invariant defined only for a compact Brillouin zone—cannot be properly formulated in this context, preventing the study of topological phases.

In this work, we present a model that is a fractional generalization of the Bernevig-Hughes-Zhang (BHZ) Hamiltonian to address the limitations of the previous model and examine the properties of its quantum geometric tensor and its quantum band topology. Moreover, we propose a tight-binding model that allows us to obtain the necessary hopping terms in real space corresponding to a given or desired dispersion.

\section{Generalized BHZ Model}
\hspace{\parindent} 
The standard BHZ model Hamiltonian \cite{Bernevig} in the orbital basis is given by
\begin{align}
\label{bhz}
    H(\mathbf{k})=&\sin k_x\sigma^x+\sin k_y\sigma^y +B(2+M-\cos k_x\nonumber\\
    &-\cos k_y)\sigma^z.
\end{align}

This model has various topological and trivial phases as the parameter $M$ is tuned. We now aim to generalize this BHZ model to satisfy the following conditions: 1.) The model must exhibit fractional dispersion. 2.) It should reduce to the standard BHZ model for integral dispersion. 3.) It must be periodic in both directions of momentum space, meaning that
\begin{align}
   H(k_x + 2\pi, k_y) =& H(k_x, k_y)\\
   H(k_x, k_y + 2\pi) =& H(k_x, k_y). 
\end{align}

This periodicity ensures a compact momentum space corresponding to the first Brillouin zone, thereby enabling a well-defined band Chern number. As a result, the topological properties of this fractional multiband system can be systematically investigated.

A model that satisfies all these conditions is
\begin{align}
\label{genbhz}
    H(\mathbf{k})=&\text{sgn}(\sin k_x)|\sin k_x|^\alpha\sigma^x+\text{sgn}(\sin k_y)|\sin k_y|^\alpha\sigma^y \nonumber\\
    &+B(2+M-\cos k_x-\cos k_y)\sigma^z.
\end{align}

This model reduces to the BHZ model for $\alpha=1$. We note further that for $M=0$, the system is gapless at $\mathbf{k}=(0,0)$ and the low energy excitations are described by the massless two-dimensional fractional Dirac Hamiltonian
\begin{align}
    H=\text{sgn}(k_x)|k_x|^\alpha\sigma^x+\text{sgn}(k_y)|k_y|^\alpha\sigma^y,
\end{align}
which was studied before \cite{Biscocho2024}. 

In the standard tight-binding or long-wavelength BHZ model, momentum terms scale with integer powers. Because the momentum vector $\mathbf{k}$ is defined across the entire Brillouin zone, its components can take on negative values \cite{Bernevig}. When the dispersion relation is generalized to a fractional exponent $\alpha$, evaluating $k_j^\alpha$ for negative values of $k_j$ results in complex numbers and multi-valued functions \cite{Laskin2018}. Such is also the case for $\sin k_j$ in Eq. \eqref{bhz}. In this formulation, the absolute value functions ensure that the model remains well-defined for values of $\sin \;k_j<0$, avoiding a non-Hermitian or multivalued Hamiltonian. The signum functions then restore the original signs lost after taking the absolute value.

The eigenfunctions of Eq. \eqref{genbhz} are
\begin{align}
\label{bloch}
    |\pm\rangle=\frac{1}{\sqrt{2d(\mathbf{k})[d(\mathbf{k})\pm d_3(\mathbf{k})]}}
\begin{pmatrix}
   d_3(\mathbf{k})\pm d(\mathbf{k})\\
    d_1(\mathbf{k})+id_2(\mathbf{k})
\end{pmatrix},
\end{align}
with the corresponding energy eigenvalues
\begin{align}
\label{energy_eigen}
E_\pm=\pm\sqrt{d_1(\mathbf{k})^2+d_2(\mathbf{k})^2+d_3(\mathbf{k})^2}\equiv \pm d(\mathbf{k}).
\end{align}
 Here, we have
 \begin{align}
\label{d123}
\begin{split}
    d_1(\textbf{k})&=\text{sgn}(\text{sin}k_x)|\text{sin}k_x|^\alpha, \\d_2(\textbf{k})&=\text{sgn}(\text{sin}k_y)|\text{sin}k_y|^\alpha , \\d_3(\textbf{k})&=B(2+M-\text{cos}k_x-\text{cos}k_y).
\end{split}
\end{align}
We will use these Bloch eigenfunctions to calculate the components of the band quantum geometric tensor.

\section{Tight-binding BHZ Model}
\hspace{\parindent} 
To connect the fractional Bloch Hamiltonian to realizable lattice systems, it is necessary to determine the real-space hopping amplitudes $t(\mathbf{r})$ corresponding to a prescribed fractional dispersion $\epsilon(\mathbf{k})$ \cite{Dartora2021}. In fractional lattice models, the standard approach is to introduce a suitable definition of fractional derivatives, such as the Gr\"{u}nwald-Letnikov derivative, and extract the hopping amplitudes from the resulting infinite series expansion \cite{Dartora2021}. While effective in simple settings, this procedure becomes cumbersome to generalize to multiband systems. Here we propose an alternative approach. Specifically, we determine the hopping amplitudes $t(\mathbf{r})$ associated with an arbitrary fractional dispersion $\varepsilon(\mathbf{k})$ directly from the fractional Bloch Hamiltonian $H(\mathbf{k})$ via the conventional Fourier transform. This construction is straightforward and readily generalizes to higher dimensions, multiband systems, and fractional models beyond the one considered in this work.

To start, we recall the general form of a tight-binding Hamiltonian
\begin{align}
\label{tight binding H}
H=\sum_{i,j,\alpha,\beta}t_{i\alpha,j\beta}c^\dagger_{i\alpha}c_{j\beta},
\end{align}
where $i$ and $j$ label the lattice sites, $\alpha$ and $\beta$ are the sublattice atoms, $c^\dagger$ and $c$ are the electron creation and annihilation operators, respectively, and $t_{i\alpha,j\beta}$ is the hopping amplitude.

The Fourier transform of the operators are
\begin{align}
\label{eq:FT operators1}
    c_{i\alpha}=&\frac{1}{\sqrt{N}}\sum_{\mathbf{k}}e^{i\mathbf{R}_i\cdot\mathbf{k}}c_{\mathbf{k}\alpha}\\
\label{eq:FT operators2}
    c_{i\alpha}^\dagger=&\frac{1}{\sqrt{N}}\sum_{\mathbf{k}}e^{-i\mathbf{R}_i\cdot\mathbf{k}}c_{\mathbf{k}\alpha}^\dagger,
\end{align}
where $\mathbf{R}_i$ is the position of the $i$th lattice site and $N$ is the number of lattice sites.

We now write $\mathbf{R}_j=\mathbf{R}_i+\mathbf{r}_{ij}$, where $\mathbf{r}_{ij}$ is a vector from $\mathbf{R}_i$ to $\mathbf{R}_j$, and assume that the hopping amplitude depends only on the relative position
\begin{align}
\label{hopping amp}
t_{i\alpha,j\beta}=t(\mathbf{R}_i,\mathbf{R}_j)_{\alpha\beta}=t(\mathbf{R}_j-\mathbf{R}_i)_{\alpha\beta}=t(\mathbf{r}_{ij})_{\alpha\beta}.
\end{align}

Substituting Eqs.\eqref{eq:FT operators1} and \eqref{eq:FT operators2} into the general tight-binding Hamiltonian in Eq. \eqref{tight binding H}, we get
\begin{align}
    H =& \frac{1}{N} \sum_{i,j,\alpha,\beta} \sum_{\mathbf{k},\mathbf{k}'} e^{-i\mathbf{R}_i\cdot\mathbf{k}} e^{i\mathbf{R}_j\cdot\mathbf{k}'} t_{i\alpha,j\beta} c_{\mathbf{k}\alpha}^\dagger c_{\mathbf{k}'\beta}.
\end{align}

% Note that we introduce two independent momentum indices, $\mathbf{k}$ and $\mathbf{k}'$, for the creation and annihilation operators. Rearranging the above equation, we get:
% \begin{align}
%     H = \frac{1}{N} \sum_{i,j,\alpha,\beta} \sum_{\mathbf{k},\mathbf{k}'} e^{-i\mathbf{R}_i\cdot\mathbf{k}} e^{i\mathbf{R}_j\cdot\mathbf{k}'} t_{i\alpha,j\beta} c_{\mathbf{k}\alpha}^\dagger c_{\mathbf{k}'\beta}.
% \end{align}

We can rewrite the exponential terms as
\begin{align}
    e^{-i\mathbf{R}_i\cdot\mathbf{k}} e^{i(\mathbf{R}_i + \mathbf{r}_{ij})\cdot\mathbf{k}'} = e^{i\mathbf{R}_i\cdot(\mathbf{k}'-\mathbf{k})} e^{i\mathbf{r}_{ij}\cdot\mathbf{k}'}
\end{align}

Since the hopping amplitude depends only on the relative position, Eq. \eqref{hopping amp}, we can separate and transform the spatial sums: the sum over $j$ transforms into a sum over the relative lattice vectors $\mathbf{r}_{ij} \rightarrow \mathbf{r}$, while the sum over $i$ becomes a sum over the absolute lattice sites $\mathbf{R}_i$. We therefore have
\begin{align}
\label{sums}
    H =& \sum_{\alpha,\beta} \sum_{\mathbf{k},\mathbf{k}'} \left( \frac{1}{N} \sum_{\mathbf{R}_i} e^{i\mathbf{R}_i\cdot(\mathbf{k}'-\mathbf{k})} \right)\nonumber\\
    &\times\left( \sum_{\mathbf{r}} t(\mathbf{r})_{\alpha\beta} e^{i\mathbf{r}\cdot\mathbf{k}'} \right) c_{\mathbf{k}\alpha}^\dagger c_{\mathbf{k}'\beta}.
\end{align}

Using 
\begin{align}
    \frac{1}{N} \sum_{\mathbf{R}_i} e^{i\mathbf{R}_i\cdot(\mathbf{k}'-\mathbf{k})} = \delta_{\mathbf{k},\mathbf{k}'},
\end{align}
the Hamiltonian then reduces to
\begin{equation}
    H = \sum_{\mathbf{k},\alpha,\beta} c_{\mathbf{k}\alpha}^\dagger H(\mathbf{k})_{\alpha\beta} c_{\mathbf{k}\beta},
    \label{eq:3.9}
\end{equation}
where the matrix elements of the Bloch Hamiltonian, $H(\mathbf{k})_{\alpha\beta}$, are related to the hopping amplitudes by
\begin{equation}
    H(\mathbf{k})_{\alpha\beta} \equiv \sum_{\mathbf{r}} t(\mathbf{r})_{\alpha\beta} e^{i\mathbf{r}\cdot\mathbf{k}}.
    \label{blochamiltonian}
\end{equation}

Inverting this then gives us
\begin{align}
\label{eq:hop}
t(\mathbf{r})_{\alpha\beta}=\sum_{\mathbf{k}}H(\mathbf{k})_{\alpha\beta}e^{-i\mathbf{r}\cdot\mathbf{k}}.
\end{align}

All of these are, of course well-known and standard. When applied to the conventional BHZ model, we get the known hopping terms
\begin{align}
    t_{11}(\mathbf{r})
    =& B(2+M)\delta_{\mathbf{r},0}-\frac{B}{2}(\delta_{\mathbf{r},\mathbf{x}}+\delta_{\mathbf{r},-\mathbf{x}})\nonumber\\
    &-\frac{B}{2}(\delta_{\mathbf{r},\mathbf{y}}+\delta_{\mathbf{r},-\mathbf{y}})\\
    t_{12}(\mathbf{r})
    =&\frac{1}{2i}(\delta_{\mathbf{r},\mathbf{x}}-\delta_{\mathbf{r},-\mathbf{x}})-\frac{1}{2}(\delta_{\mathbf{r},\mathbf{y}}-\delta_{\mathbf{r},-\mathbf{y}})\\
    t_{21}(\mathbf{r})
    =&\frac{1}{2i}(\delta_{\mathbf{r},\mathbf{x}}-\delta_{\mathbf{r},-\mathbf{x}})+\frac{1}{2}(\delta_{\mathbf{r},\mathbf{y}}-\delta_{\mathbf{r},-\mathbf{y}})\\
    t_{22}(\mathbf{r})
    =& -B(2+M)\delta_{\mathbf{r},0}+\frac{B}{2}(\delta_{\mathbf{r},\mathbf{x}}+\delta_{\mathbf{r},-\mathbf{x}})\nonumber\\
    &+\frac{B}{2}(\delta_{\mathbf{r},\mathbf{y}}+\delta_{\mathbf{r},-\mathbf{y}}).
\label{bhz hop}
\end{align}

\subsection{Fractional 1D Hamiltonian} 
\label{benchmark}
We now show that the method reviewed above extends naturally to fractional systems, in contrast to the more elaborate approaches typically employed in applications of fractional quantum mechanics to lattice models. We use Eq. \eqref{eq:hop} to the one-dimensional fractional Hamiltonian
\begin{align}
\label{eq:fractionalH}
    H=-\frac{\hbar^\alpha}{2m_\alpha}\frac{d^\alpha }{dx^\alpha},
\end{align}
and show that we get the same result for the hopping amplitude
\begin{align}
    t(r)=(-1)^r\frac{\Gamma(\alpha+1)}{\Gamma(\alpha-r)(r+1)!}t_0,
\label{t_r}
\end{align}
obtained by using the Gr\"{u}nwald-Letnikov definition of fractional derivative \cite{Dartora2021}.

The momentum representation of Eq. \eqref{eq:fractionalH} can be interpreted as low-energy effective Bloch Hamiltonian
\begin{align}
    H(k)=\frac{\hbar^\alpha k^\alpha}{2m_\alpha}.
\end{align}
\noindent Its associated hopping matrix in $k$ space is
\begin{align}
    \hat{T}(k)=(2-\alpha)t_0+\frac{\hbar^\alpha}{2m_\alpha}|k|^\alpha.
    \label{hopping Dart}
\end{align}

Performing an inverse Fourier transform on this matrix gives us,
\begin{align}
t(r)=&\frac{1}{2\pi}\int^{\infty}_{-\infty}\hat{T}(k)e^{ikr}dk\\=&\frac{1}{2\pi}\int^{\infty}_{-\infty}\left[(2-\alpha)t_0+\frac{\hbar^\alpha}{2m^\alpha}|k|^\alpha]e^{ikr}\right]dk\\
    =&\sqrt{2\pi}(2-\alpha)t_0\delta(r)\nonumber\\
    &+\frac{1}{\sqrt{2\pi}}\frac{\hbar^\alpha}{2m_\alpha}\int_0^{\infty}k^\alpha(e^{ikr}+e^{-ikr})dk\\
    =&\sqrt{2\pi}(2-\alpha)t_0\delta(r)\nonumber\\
    &+\frac{1}{\sqrt{2\pi}}\frac{\hbar^\alpha}{2m_\alpha}\int^\infty_02k^\alpha\cos (kr)dk\\
    =&\sqrt{2\pi}(2-\alpha)t_0\delta(r)-\sqrt{\frac{2}{\pi}}\frac{\hbar^\alpha}{2m_\alpha}\sin\left(\frac{\pi\alpha}{2}\right)\nonumber\\
    &\times\Gamma(\alpha+1)|r|^{-(\alpha+1)}.
    \label{eq 3.24}
\end{align}

Next, we rewrite Eq. \eqref{t_r} and show that it is proportional to Eq. \eqref{eq 3.24}. Using Euler's reflection formula, 
\begin{align}
    \Gamma(\alpha)\Gamma(1-\alpha)=\frac{\pi}{\text{sin}(\pi\alpha)},
\end{align}
\noindent we can rewrite the factor $\Gamma(\alpha-r)$ in Eq. \eqref{t_r} as 
\begin{align}
    \Gamma(\alpha-r)=&\Gamma(1-(r-\alpha+1))\\
    =&\frac{\pi}{\text{sin}(\pi(r-\alpha+1))\Gamma(r-\alpha+1)}.
    \label{euler}
\end{align}

Since we are interested in the long-wavelength limit, that is, large $r$, we can use Stirling's approximation
\begin{align}
     \Gamma(z)\approx z^z e^{-z}\sqrt{2\pi z}
\end{align}
\noindent for the factors $(r+1)!$ and $\Gamma(r-\alpha+1)$ in Eqs. \eqref{t_r} and \eqref{euler}, respectively, to get
\begin{align}
    (r+1)!=\Gamma(r+2)\approx&(r+2)^{(r+2)}e^{-(r+2)}\nonumber\\
    &\times\sqrt{2\pi(r+2)}\\
   \Gamma(r-\alpha+1)\approx&(r-\alpha+1)^{(r-\alpha+1)}e^{-(r-\alpha+1)}\nonumber\\
   &\times\sqrt{2\pi(r-\alpha+1)}.
\end{align}

Using these, Eq. \eqref{t_r} can now be written as
\begin{align}
    t(r)=&(-1)^r\frac{\Gamma(\alpha+1)}{\Gamma(1-(r-\alpha+1))(r+1)!}t_0\\ 
    =&(-1)^r\frac{\Gamma(\alpha+1) \text{sin}(\pi(r-\alpha+1)) \Gamma(r-\alpha+1)}{\pi\Gamma(r+2)}t_0\\
    \approx&(-1)^{r+1}\Gamma(\alpha+1)\text{sin}(\pi(r-\alpha))e^{\alpha-3}\nonumber\\
    &\times\sqrt{\frac{r-\alpha+1}{r+2}}\frac{(r-\alpha+1)^{r-\alpha+1}}{(r+2)^{r+2}}.
\end{align}

For large $r$, this gives
\begin{align}
    t(r)\sim\Gamma(\alpha+1)r^{-(\alpha+1)},
\end{align}
\noindent which exhibits the same $\Gamma(\alpha+1) r^{-(\alpha+1)}$ dependence as in Eq. \eqref{eq 3.24}. Our method, however, has the advantage that it can be easily used for the general case of higher dimensions and multiband systems with fractional dispersion.

\subsection{The Behavior of Hopping under Fractional Tuning}

Having established how to calculate the hopping amplitudes from the momentum-space Hamiltonian and having shown that they hold for fractional systems, we now apply the same formulation to our generalized BHZ model in Eq. \eqref{genbhz}.
 
\begin{figure*}[hbt]
    \centering
    \includegraphics[width=\textwidth]{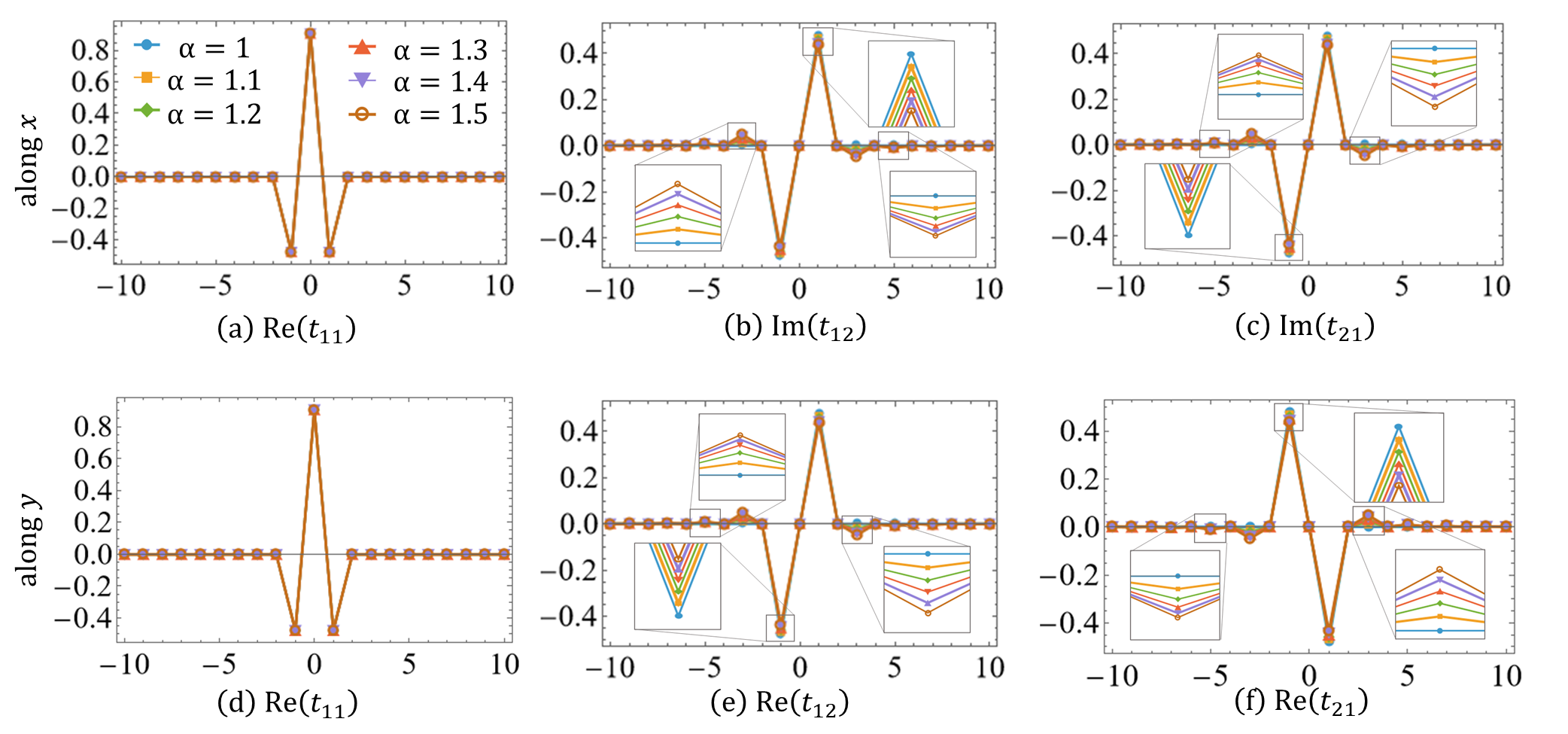} 
    \caption{Components of the $x$-dependent hopping terms: (a) $\text{Re}(t_{11})$, (b) $\text{Im}(t_{12})$, and (c) $\text{Im}(t_{21})$, and of the $y$-dependent hopping terms: (d) $\text{Re}(t_{11})$, (e) $\text{Re}(t_{12})$, and (f) $\text{Re}(t_{21})$ of the generalized BHZ model for dispersions $\alpha = 1 \text{ to } 1.5$ in increments of 0.1.}
    \label{fig:Hoppings}
\end{figure*}

Figs. \ref{fig:Hoppings}(a) to (f) show the real and imaginary parts of the hopping along $x$ and $y$ directions. Figs. \ref{fig:Hoppings}(a) and (d) show the real part of the intraorbital hopping term $t_{11}$. It shows that it is unaffected when the dispersion is tuned to fractional values. Its imaginary part is identically zero. We also obtained the same result for $t_{22}$. 

In contrast, both the real and imaginary parts of the interorbital hoppings, Figs. \ref{fig:Hoppings}(b), (c), (e), and (f), change when the dispersion is tuned to fractional values. Here, we fix $M=1$ as the hopping terms are independent of $M$. In the BHZ model, these terms play the role of an effective spin-orbit coupling that drives band inversion and determines the topological character of the system \cite{Bernevig}. Physically, they encode the phase acquired by an electron as it hops between orbitals. As shown in Fig.~\ref{fig:Hoppings}(b) and (c), the imaginary hopping amplitudes exhibit odd-parity symmetry, resulting in an antisymmetric profile about the central lattice site $(r=0)$. For fractional $\alpha$, nonzero imaginary hopping amplitudes emerge beyond nearest neighbors; in particular, all odd $n$th-nearest-neighbor hoppings acquire finite imaginary components. Such long-range hopping terms are a characteristic feature of systems with fractional dispersion \cite{Stickler2013}.

Note that we can express the hopping terms $t$ either in terms of their imaginary and real components, $t=t_R+it_I$, or in terms of their magnitude and phase, $t=|t|e^{i\theta}$. Here, the phase $\theta$ is related to the imaginary and real components by the relation
\begin{align}
\tan\theta=\frac{t_I}{t_R}.
\label{tan}
\end{align}

Since the imaginary components of the diagonal elements $t_{11}$ and $t_{22}$, are all zero for different values of $\alpha$, then their corresponding phases are also zero for $t_R>0$. On the other hand, the off-diagonal hopping terms acquire a phase $\pm\frac{\pi}{2}$ for every odd nearest-neighbor in the fractional cases, and a zero phase for the $\alpha=1$ case. This can be seen using Eq. \eqref{tan}, along with Figs. \ref{fig:Hoppings} (b), (c), (e), and (f), where $\mathcal{I}m(t_{12})$ and $\mathcal{I}m(t_{21})$ are nonzero but $\mathcal{R}e(t_{12})=\mathcal{R}e(t_{21})=0$. Note, however, that for all values of $\alpha$, the phases gained from the nearest-neighbor hoppings are all $\pm\frac{\pi}{2}$.

In the $y$-direction, the behavior is reversed.  The real components of the off-diagonal elements, $\text{Re}(t_{12})$ and $\text{Re}(t_{21})$, as shown in Figs. \ref{fig:BC QM}(e) and (f) are nonzero and vary as the dispersion is tuned to fractional values. Instead, all imaginary components of the hopping terms are exactly zero, which implies that any hopping in the $y$-direction does not acquire a phase.

\section{Quantum Geometric Tensor Under Fractional Dispersion}\label{frac QGT}
\hspace{\parindent}
Having shown how to construct the corresponding tight-binding model for our fractional BHZ model, we now investigate the effects of the fractional dispersion on the band quantum geometry and topology. First, recall that the Berry curvature and the quantum metric of the $n$th band can be computed using the following convenient formulas
\begin{align}
\label{eq:BC}
    F^{(n)}_{xy}(\mathbf{k})=-2\mathcal{I}m\sum_{m\neq n}\frac{\langle n|\partial_xH|m\rangle\langle m|\partial_yH|n\rangle}{(E_n-E_m)^2}
\end{align}
and
\begin{align}
\label{eq:QM}
    g^{(n)}_{ij}(\mathbf{k})=\mathcal{R}e\sum_{m\neq n}\frac{\langle n|\partial_iH|m\rangle\langle m|\partial_jH|n\rangle}{(E_n-E_m)^2}.
\end{align}

Here, $|m\rangle$ is the cell-periodic Bloch eigenket of the $m$th band, $E_m$ is the corresponding energy eigenvalue, and $\partial_i\equiv\partial/\partial k^i$.

\begin{figure*}[hbt]
    \centering
    \includegraphics[width=\textwidth]{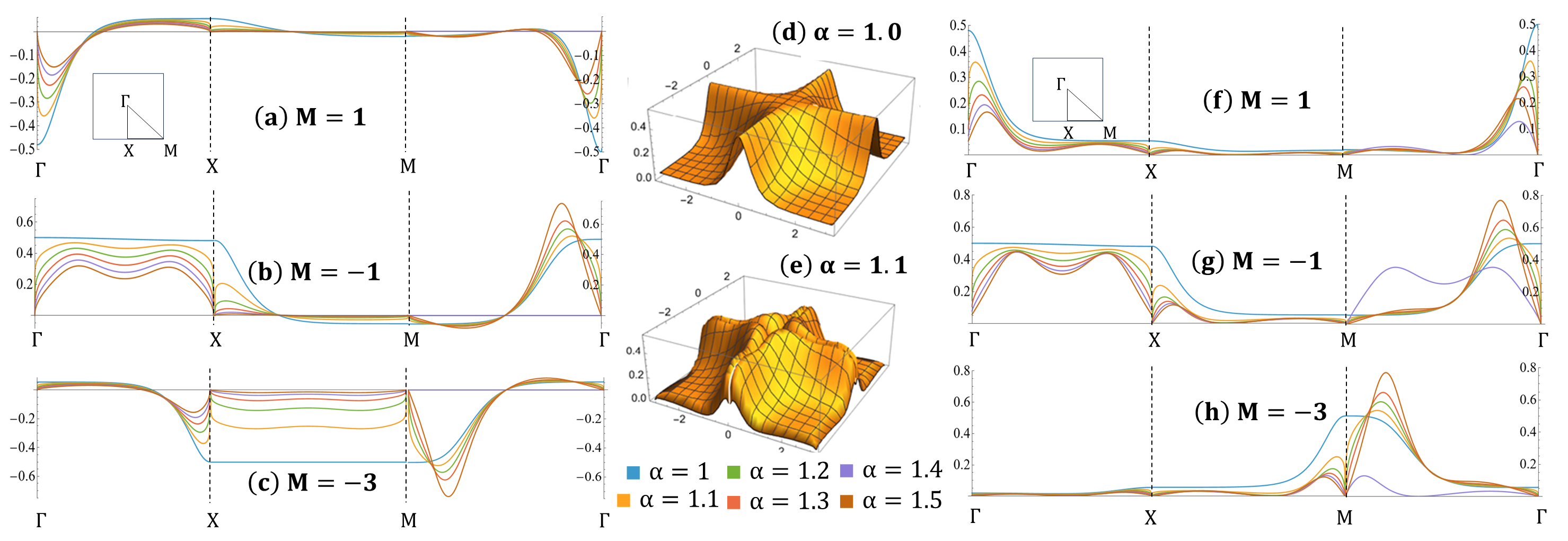} 
    \caption{Berry curvature plots in the first Brillouin zone for the BHZ model with integer dispersions $\alpha = 1 \text{ to } 1.5$ in increments of 0.1, shown for (a) $M = 1$, (b) $M=-1$, and (c) $M=-3$  with $B = 1$ fixed., 3D plots of the Berry curvature of the upper band for $M=-1$ and dispersion (d) $\alpha=1$ and (e) $\alpha=1.1$, and Trace of the quantum metric plots in the first Brillouin zone for the BHZ model with integer dispersions $\alpha = 1 \text{ to } 1.5$ in increments of 0.1, shown for (f) $M = 1$, (g) $M=-1$, and (h) $M=-3$  with $B = 1$ fixed.}
    \label{fig:BC QM}
\end{figure*}

We use these formulas, along with the eigenfunctions in Eq. \eqref{bloch} to calculate the components of the QGT for our generalized BHZ model given by Eq. \eqref{genbhz}. We plot the Berry curvature and quantum metric along the high-symmetry line path $\Gamma\rightarrow X \rightarrow M\rightarrow\Gamma$ in Figs. \ref{fig:BC QM}(a) to (c) and (f) to (h), respectively. 

 Since the Berry curvature of the lower band is simply the negative of that of the upper band, it is sufficient to focus our analysis on the latter. The Berry curvatures of the upper band are plotted in Figs. \ref{fig:BC QM}(a) to (c) for $\alpha = 1$ to $1.5$ in increments of $0.1$, and values of $M$ at different topological phases. We restrict our discussion to $\alpha \geq 1$, as values of $\alpha < 1$ are generally considered unphysical. This is evident in the continuum limit, where $\alpha < 1$ leads to divergences and discontinuities in physical quantities such as the Fermi velocity and the Wannier center \cite{Biscocho2024}. In the context of fractional quantum mechanics, $\alpha < 1$ also results in divergent moments, further supporting the restriction \cite{Laskin2018}.
 
The blue plots show the Berry curvatures for $\alpha = 1$, serving as a benchmark case, since our model should reproduce the well-known results of the BHZ model in this limit. Recall that for the BHZ model, $M=1$ is a trivial insulator. In Fig. \ref{fig:BC QM}(a), we observe a negative peak centered at $\Gamma$ (i.e., $\mathbf{k} = 0$), which decreases in magnitude as we tune the dispersion $\alpha$ to higher fractional values. Fig. \ref{fig:BC QM}(b) displays the Berry curvature at $M = -1$, where it is generally positive throughout the Brillouin zone. This behavior is consistent with the system being in a nontrivial topological phase with Chern number $C = +1$. 

Similarly, in Fig.~\ref{fig:BC QM}(c), corresponding to $M = -3$, the Berry curvature is generally negative across the entire Brillouin zone, consistent with the system entering a topological phase characterized by a Chern number $C = -1$.

Comparing the plots of the fractional dispersion values to our benchmark $\alpha = 1$ case, we observe that the Berry curvature is redistributed across the Brillouin zone. In particular,  our results show that the single peaks observed in the corresponding BHZ plots, shown in Fig. \ref{fig:BC QM}(d), are weakened in magnitude and fold inward to their points of symmetry, splitting the peaks into four localized clusters (as we still have a four-fold rotational symmetry), as shown in Fig. \ref{fig:BC QM}(e).

Despite these shape deformations, the overall sign of the Berry curvature across the Brillouin zone remains the same. For instance,  the signs of the Berry curvature for the fractional cases in Fig. \ref{fig:BC QM}(a) remain mostly negative, just as the $\alpha=1$ case. The same can be observed for the fractional and $\alpha=1$ cases in Figs. \ref{fig:BC QM}(b) and (c). Because of this, we ask if the Chern number is unaffected when $\alpha$ is changed to fractional values. We will address this question in the next subsection.

As with the Berry curvature, we observe a redistribution of the quantum metric across the Brillouin zone, as shown in Figs. \ref{fig:BC QM}(f) to (h). Similarly, each peak in the standard BHZ case splits into four clustered peaks with weakened strengths when the dispersion is modified. These plots demonstrate that tuning the dispersion from the integer value $\alpha = 1$ (corresponding to the BHZ model) to a fractional value, such as $\alpha = 1.1 \text{ to } 1.5$ as shown, generally alters the quantum distance between Bloch states throughout the Brillouin zone. In two-dimensional systems with integer dispersion, the trace of the quantum metric equals the absolute value of the Berry curvature. A comparison of Figs. \ref{fig:BC QM}(a) to (c) and (f) to (h) reveals that the Berry curvature and the trace of the quantum metric exhibit similar behavior even for fractional values of $\alpha$, providing numerical evidence that this relationship may extend to general fractional systems.

In summary, tuning the dispersion parameter to fractional values significantly redistributes the components of the QGT across momentum space. Local geometric features that appear as single, distinct peaks in the standard integer-dispersion case $(\alpha=1)$ deform and split into localized clusters of four weakened peaks under fractional tuning $(\text{e.g., }\alpha=1.1)$. This behavior is observed in both the Berry curvature and the trace of the quantum metric. Despite these structural deformations, the overall sign distribution of the Berry curvature remains consistent with the $\alpha=1$ case. 

\subsection{Invariance of the Chern Number}
\hspace{\parindent} 
As discussed above, tuning the dispersion exponent to fractional values generally redistributes both the quantum metric and the Berry curvature. We now examine whether such tuning also modifies the Chern number, which is a topological invariant characterizing the band structure. In particular, we investigate whether varying the dispersion exponent away from integer values can induce topological phase transitions. Unlike the fractional Dirac model, the generalized BHZ model possesses a compact Brillouin zone, which allows the Chern number to be computed unambiguously by integrating the Berry curvature over the first Brillouin zone. We therefore evaluate the Chern number of the generalized BHZ model through a direct integration of the Berry curvature over the first Brillouin zone
\begin{equation}
\label{eq:chernumber}
C_n = \frac{1}{2\pi} \iint_{\text{BZ}} F_{xy}^{(n)}(\mathbf{k}) \, d^2k.
\end{equation}

\begin{figure}[hbt]
    \centering
    \includegraphics[width=0.9\linewidth]{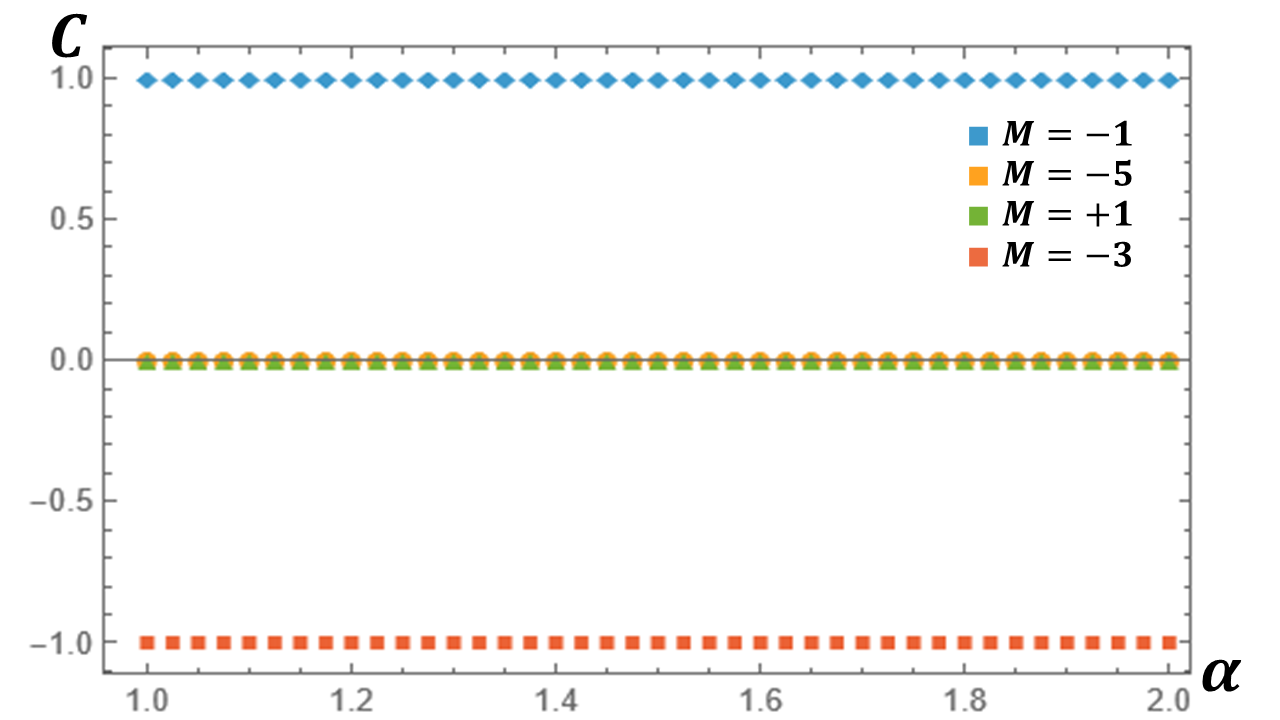}
    \caption{Chern number values for the generalized BHZ model with $B = 1$ fixed, shown for different values of $\alpha$. Values of M are chosen such that each value represents different phases.}
    \label{fig: CN}
\end{figure}

Figure~\ref{fig: CN} shows the Chern number as a function of $\alpha$, where $\alpha$ is varied from $\alpha=1.0$, corresponding to the conventional BHZ model, to fractional values up to $\alpha=2.0$, for several values of the parameter $M$. Although tuning $\alpha$ redistributes the Berry curvature throughout the Brillouin zone, the Chern number remains unchanged. This indicates that, for fixed $M$, the system remains in its respective topological phase despite variations in $\alpha$.

This demonstrates that the robustness of the system’s global topological phase is preserved even as the local geometric structure and effective lattice connectivity are continuously modified through the fractional dispersion parameter. The invariance of the Chern number under fractional tuning further underscores the distinction between local geometry and global topology. In particular, tuning the dispersion exponent to fractional values redistributes the Berry curvature across the Brillouin zone while leaving its total integral, and hence the Chern number, unchanged.

\section{Generalization}
Having analyzed the quantum geometry and topology of the fractional generalization of the BHZ model as a concrete example, we now present more general arguments showing that tuning the dispersion exponent to fractional values does not alter the Chern number. Furthermore, for two-band systems, the fundamental relation between the quantum metric and the Berry curvature,
\begin{align}
\det g_{\mu\nu}=\frac{1}{4}|\Omega|^2
\end{align}
remains preserved under fractional tuning of the dispersion.

We start with the Bloch Hamiltonian
\begin{align}
    H(\mathbf{k})=\mathbf{d}\cdot\boldsymbol{\sigma}
\end{align}
and assume that the vector $\mathbf{d}$ is a function of $\mathbf{k}$ via
\begin{align}
   \mathbf{d}=\mathbf{d}(s^\alpha_\mu,c^\alpha_\mu,\sin k_\mu,\cos k_\mu)
\end{align}
where
\begin{align}
    s^\alpha_\mu:=&\sgn (\sin k_\mu)|\sin k_\mu|^\alpha\\
    c^\alpha_\mu:=&\sgn (\cos k_\mu)|\cos k_\mu|^\alpha.
\end{align}

Just like in the generalization of the BHZ model, this dependence ensures that there is periodicity in $\mathbf{k}$ space and that the Hamiltonian is Hermitian and single-valued even with fractional exponents.

We now consider a coordinate transformation
\begin{align}
\label{eq:generalcoordinatetransform}
    p^a=p^a(s^\alpha_\mu,c^\alpha_\mu,\sin k_\mu,\cos k_\mu)
\end{align}
in $\mathbf{k}$ space, with the requirement that in this new coordinates, the quantum metric and Berry curvature do not have explicit dependence on $\alpha$. An example will clarify what we mean. Consider the simplified fractional BHZ
\begin{align}
\label{eq:genbhzsimplified}
    H(\mathbf{k})=&\text{sgn}(\sin k_x)|\sin k_x|^\alpha\sigma^x+\text{sgn}(\sin k_y)|\sin k_y|^\alpha\sigma^y \nonumber\\
    &+M\sigma^z.
\end{align}
Then the momentum-space coordinate transformation
\begin{align}
    \sin p^x:=&\text{sgn}(\sin k_x)|\sin k_x|^\alpha\\
    \sin p^y:=&\text{sgn}(\sin k_y)|\sin k_y|^\alpha
\end{align}
transforms the Hamiltonian into
\begin{align}
\label{eq:genbhzsimplified2}
    H(\mathbf{k})=\sin p_x\sigma^x+\sin p_y\sigma^y+M\sigma^z.
\end{align}
The QGT calculated from the eigenstates of this Hamiltonian is therefore independent of $\alpha$. Note, however, that in general more than one overlapping coordinate patch may be required to fully cover the Brillouin zone. Determining such a set of coordinates is itself nontrivial in general. One possible approach is the following. First, compute the quantum metric from the original Hamiltonian with explicit $\alpha$ dependence, expressed for example in $\mathbf{k}$ coordinates. One may then diagonalize the quantum metric and identify the appropriate new coordinates $\mathbf{p}$ as
\begin{align}
    (dp^x)^2=g_{xx}(\alpha)(dk^x)^2,\;\;\;(dp^y)^2=g_{yy}(\alpha)(dk^y)^2.
\end{align}
In this new coordinates, $\alpha$ does not appear explicitly.

We now use the following formula for the quantum metric for two-band systems
\begin{align}
\label{eq:quantummetrictwoband}
    g^{(\pm)}_{\mu\nu}=\frac{1}{4d^2}[\partial_\mu\mathbf{d}\cdot\partial_\nu\mathbf{d}-\frac{1}{d^2}(\partial_\mu\mathbf{d}\cdot\mathbf{d})(\partial_\nu\mathbf{d}\cdot\mathbf{d})].
\end{align}

Under coordinate transformation
\begin{align}
\partial_\mu d_j=\frac{\partial p^a}{\partial k^\mu}\frac{\partial d_j}{\partial p^a}
\end{align}
so that Eq. \eqref{eq:quantummetrictwoband} can be written as
\begin{align}
\label{eq:quantummetrictwoband2}
    g^{(\pm)}_{\mu\nu}=\frac{\partial p^a}{\partial k^\mu}\frac{\partial p^b}{\partial k^\nu}\Tilde{g}^{(\pm)}_{ab},
\end{align}
where $\Tilde{g}^{(\pm)}_{ab}$ is the transformed metric given by
\begin{align}
   \Tilde{g}^{(\pm)}_{ab}=\frac{1}{4d^2}[\partial_a\mathbf{d}\cdot\partial_b\mathbf{d}-\frac{1}{d^2}(\partial_a\mathbf{d}\cdot\mathbf{d})(\partial_b\mathbf{d}\cdot\mathbf{d})].
\end{align}

Similarly, for the Berry curvature, we have
\begin{align}
\label{eq:berrycurvaturetransform}
    F^{(\pm)}_{\mu\nu}=\pm\frac{\partial p^a}{\partial k^\mu}\frac{\partial p^b}{\partial k^\nu} \tilde{F}^{(\pm)}_{ab}.
\end{align}

Under the change of coordinates, $\mathbf{k}\rightarrow\mathbf{p}$, the Chern number Eq. \eqref{eq:chernumber} now becomes
\begin{align}
\label{eq:chernk}
    C_n =& \frac{1}{2\pi} \int_{\text{BZ}}d^2k F_{\mu\nu}^{(n)}(\mathbf{k})\\
    \label{eq:chernchanged}
    =& \frac{1}{2\pi} \int_{\text{BZ}}d^2pJ \frac{\partial p^a}{\partial k^\mu}\frac{\partial p^b}{\partial k^\nu}\tilde{F}_{ab}^{(n)}(\mathbf{k}),
\end{align}
where $J$ is the Jacobian of the transformation:
\begin{align}
    J:=\left|\frac{\partial(k^x,k^y)}{\partial(p^1,p^2)}\right|=\frac{\partial k^x}{\partial p^1}\frac{\partial k^y}{\partial p^2}-\frac{\partial k^y}{\partial p^1}\frac{\partial k^x}{\partial p^2}.
\end{align}

We can expand the integrand in Eq. \eqref{eq:chernchanged} explicitly
\begin{align}
    C_n =& \frac{1}{2\pi} \int_{\text{BZ}}d^2pJ \frac{\partial p^a}{\partial k^\mu}\frac{\partial p^b}{\partial k^\nu}\tilde{F}_{ab}^{(n)}(\mathbf{k})\\
    =& \frac{1}{2\pi} \int_{\text{BZ}}d^2pJ \left(\frac{\partial p^1}{\partial k^x}\frac{\partial p^2}{\partial k^y}-\frac{\partial p^2}{\partial k^x}\frac{\partial p^1}{\partial k^y}\right)\tilde{F}_{12}^{(n)}(\mathbf{p}).
\end{align}
The factor inside the parenthesis is just the inverse of the jacobian,
\begin{align}
    J^{-1}=\left(\frac{\partial p^1}{\partial k^x}\frac{\partial p^2}{\partial k^y}-\frac{\partial p^2}{\partial k^x}\frac{\partial p^1}{\partial k^y}\right),
\end{align}
giving
\begin{align}
\label{eq:chernp}
    C_n 
    = \frac{1}{2\pi} \int_{\text{BZ}}d^2p\tilde{F}_{12}^{(n)}(\mathbf{p}).
\end{align}

The cancellation between the Jacobian and its inverse ensures that the Chern number takes the same form in different Brillouin-zone coordinate systems, Eqs. \eqref{eq:chernk} and \eqref{eq:chernp}. In the argument above, we began with the fractional BHZ model. The same reasoning can also be viewed in reverse: starting from the conventional BHZ model, one may continuously deform the dispersion exponent $\alpha$ to fractional values. This deformation can be interpreted as a coordinate transformation in momentum space, $\mathbf{k}$, which leaves the Chern number invariant as we have shown above.

Using similar method, we can also show that the relation
\begin{align}
\label{eq:metrictoberry}
    \det g=\frac{1}{4}|F|^2,
\end{align}
which is true for two-band systems with integer dispersion, is also true for systems with fractional dispersions. To do this, we take the determinant of Eq. \eqref{eq:quantummetrictwoband2} to get
\begin{align}
\label{eq:metricdettransform}
    \det g^{(\pm)}_{\mu\nu}=(J^{-1})^2\det\tilde{g}^{(\pm)}_{ab}.
\end{align}

Similarly, from Eq. \eqref{eq:berrycurvaturetransform}
\begin{align}
\label{eq:berrycurvaturetransform2}
    F^{(\pm)}_{12}=\pm J^{-1} \tilde{F}^{(\pm)}_{12},
\end{align}
which gives
\begin{align}
\label{eq:berrycurvaturetransform3}
    |F^{(\pm)}_{12}|=(J^{-1})^2 |\tilde{F}^{(\pm)}_{12}|.
\end{align}

Hence, starting from any two-band system with integer dispersion for which Eq.~\eqref{eq:metrictoberry} holds, one may continuously deform the dispersion to fractional values. Applying Eqs.~\eqref{eq:metricdettransform} and \eqref{eq:berrycurvaturetransform3} then yields
\begin{align}
\label{eq:metrictoberry2}
\det \tilde{g}=\frac{1}{4}|\tilde{F}|^2.
\end{align}
Therefore, the fundamental relation between the determinant of the quantum metric and the Berry curvature remains preserved under fractional deformation of the dispersion.

\section{Conclusions}
We have investigated the impact of fractional dispersion on quantum band geometry by extending the BHZ model to incorporate fractional band dispersion, thereby overcoming the limitations of low-energy continuum descriptions. The resulting fractional BHZ model continuously recovers the conventional integer-dispersion limit and provides a controlled framework for studying how fractional scaling modifies quantum geometric properties.

We showed that the real-space hopping amplitudes can be obtained through a standard inverse Fourier transformation even in fractional systems. Unlike approaches commonly used in fractional quantum mechanics applied to condensed matter systems, this method extends naturally to higher-dimensional momentum spaces and multiband models. Applied to the fractional BHZ model, it reveals pronounced anisotropic hopping behavior under tuning of the dispersion parameter $\alpha$, together with the emergence of long-range hopping processes for noninteger $\alpha$.

We further found that fractional dispersion strongly redistributes the quantum geometric tensor throughout momentum space. Despite these local geometric deformations, the sign structure of the Berry curvature remains unchanged, and the relation between the trace of the quantum metric and the magnitude of the Berry curvature is preserved across the Brillouin zone.

Finally, these local geometric modifications leave the global topology intact: the Chern number remains invariant for all fractional values of $\alpha$. This demonstrates a clear distinction between tunable local quantum geometry and robust topological invariants in this exotic fractional system. Our results therefore establish a general framework for constructing tight-binding lattice models that realize multiband systems with fractional dispersion and for studying its band geometric and topological properties.

\bibliography{apssamp}% Produces the bibliography via BibTeX.

\end{document}